\documentclass[twocolumn,superscriptaddress,showpacs,prl]{revtex4}
\usepackage{mathrsfs}
\usepackage{amssymb, amsbsy, amsmath, latexsym, dsfont, array, layout, graphicx,mathrsfs,color}

\begin{document}

\title{Observation of quasiperiodic dynamics in a one-dimensional quantum walk of single photons in space}
\author{Peng Xue\footnote{gnep.eux@gmail.com}
} \affiliation{Department of Physics, Southeast University, Nanjing
211189, China}
\affiliation{State Key Laboratory of Precision
Spectroscopy, East China Normal University, Shanghai 200062, China}
\author{Hao Qin}
\affiliation{Department of Physics, Southeast University, Nanjing
211189, China}
\author{Bao Tang}
\affiliation{Department of Physics, Southeast University, Nanjing
211189, China}
\author{Barry C. Sanders}
\affiliation{Hefei National Laboratory for Physical Sciences at
Microscale and Department of Modern Physics, University of Science
and Technology of China, Hefei, 230026, China}
\affiliation{Institute for Quantum Science and Technology,
University of Calgary,
        Alberta T2N 1N4, Canada}

\begin{abstract}
We realize quasi-periodic dynamics of a quantum walker over 2.5
quasi-periods by realizing the walker as a single photon passing
through a quantum-walk optical-interferometer network. We introduce
fully controllable polarization-independent phase shifters in each
optical path to realize arbitrary site-dependent phase shifts, and
we employ large clear-aperture beam displacers, while maintaining
high-visibility interference, to enable reaching 10 quantum-walk
steps. By varying the half-wave-plate setting, we control the
quantum-coin bias thereby observing a transition from quasi-periodic
dynamics to ballistic diffusion.
\end{abstract}


\maketitle

The quantum walk (QW)~\cite{BMK+99,BFL+10} is a quantized version of
the ubiquitous random walk (RW) used for describing
diffusion~\cite{Amb03}, for probabilistic algorithms~\cite{CCD+03}
to solve constraint satisfaction problems in computer
science~\cite{Sch99}, for quantum transport in complex
systems~\cite{OPR06} and for demonstrating intriguing nonlinear
dynamical quantum phenomena~\cite{WLK+04}. In its basic formulation
a walker moves on an integer lattice with its periodic spatial sites
labelled by integers~$x$. This one-dimensional line is regarded as
being arbitrarily long and the walker commences at the origin $x=0$.
The walker carries a two-sided coin with the two sides labelled
by~$c\in\{0,1\}$.

The walker's evolution proceeds as follows: the walker flips the
coin to obtain an outcome~$c$ and then makes one step in the
positive or negative direction on the line if the coin flip yields
the outcome~$0$ or~$1$, respectively. The randomness of the coin
flip leads to a diffusion rate that increases as~$\sqrt{t}$ with~$t$
the time of evolution (which we treat as a non-negative integer
incrementing by one unit per step from $t=0$); this square-root
dependence is characteristic of diffusive spreading, with ``spread''
signifying the width of the position distribution~$P_t(x)$ at
time~$t$.

The QW eliminates random evolution by trading the coin for a quantum
two-level system, which, in our case, is the polarization state of a
single photon: horizontal~(H) or vertical~(V). Furthermore the QW
employs unitary dynamics by which the walker's position is entangled
with the coin state. When the evolution is complete, the coin state
is ignored, which mathematically corresponds to tracing out this
degree of freedom.

For this evolution, the walker's position distribution then spreads
proportional to~$t$ rather than to~$\sqrt{t}$. Proportionality of
position spreading to~$t$ is reminiscent of constant-velocity
deterministic motion and is thus known as ``ballistic transport''.
QWs are widely studied experimentally and theoretically because of
their applications to quantum algorithms~\cite{Amb03} and quantum
transport~\cite{OPR06} and also for intriguing foundational studies
of quantum phenomena such as defect-induced
localization~\cite{COR+13,SCP+11} and quantum
resonances~\cite{GAS+13,CRW+13}.

Mathematically each evolutionary ``step'' given by unitary
operator~$U$ on the joint walker-coin system is achieved by
repeating two sequential unitary operations. The first operation is
the unitary analogue of the coin flip achieved by transforming the
coin states $|c=0\rangle$ and $|c=1\rangle$ to
$\cos\theta|0\rangle+\sin\theta|1\rangle$ and
$\sin\theta|0\rangle-\cos\theta|1\rangle$, respectively,
with~$\theta$ the coin-bias parameter. The second (entangling)
operation translates the walker's position dependent on the coin
state:
$\sum_{c=0}^1\sum_{x=-\infty}^\infty\text{e}^{\text{i}\phi(x)}|x+(-1)^c\rangle\langle
x|\otimes |c\rangle\langle c|$. Typically $\phi(x)\equiv 0$ in
theory~\cite{Whaley02,BCA03,KRBE10,CG06,B06} and
experiment~\cite{KFC+09,PO10,PS08,SS12,S10,SSV+12,D05,Z07,KBF+12},
but controllably varying the phase enables the realizations of
remarkable quantum-walk properties such as quantum recurrences in
diffusion, which arise in the ``harmonic case'' $\phi(x)=2\pi xy$
for $y=q/p$; we focus on $q$, $p$ co-prime integers~\cite{WLK+04}.

The walker is manifested experimentally as a single photon and the
lattice as a set of allowed paths for the photon. Pairs of $800$nm
photons are generated via type-I spontaneous parametric down
conversion (SPDC) in two $0.5$mm-thick
nonlinear-$\beta$-barium-borate (BBO) crystals cut at $29.41^\circ$,
which are pumped by a $100$mW $400$nm continuous-wave diode laser.
Triggering on one photon prepares the other photon in the pair as a
single-photon state. After spectral filtering with
full-width-at-half-maximum (FWHM) $3$nm, individual down-converted
photons are steered into the optical modes of the linear-optical
network formed by a series of birefringent calcite beam displacers
(BDs), half-wave plates (HWPs) and phase shifters (PSs). Output
photons are detected using avalanche photo-diodes (APDs, $7$ns time
window) with a dark count rate of less than 100s$^{-1}$ whose
coincident signals---monitored using a commercially available
counting logic---are used to postselect two single-photon events.
The total coincident counts are around $20,000$ over $40$s. The
probability of creating more than one photon pair is less than
$10^{-4}$ and can thus be neglected.

\begin{figure}
\includegraphics[width=8.5cm]{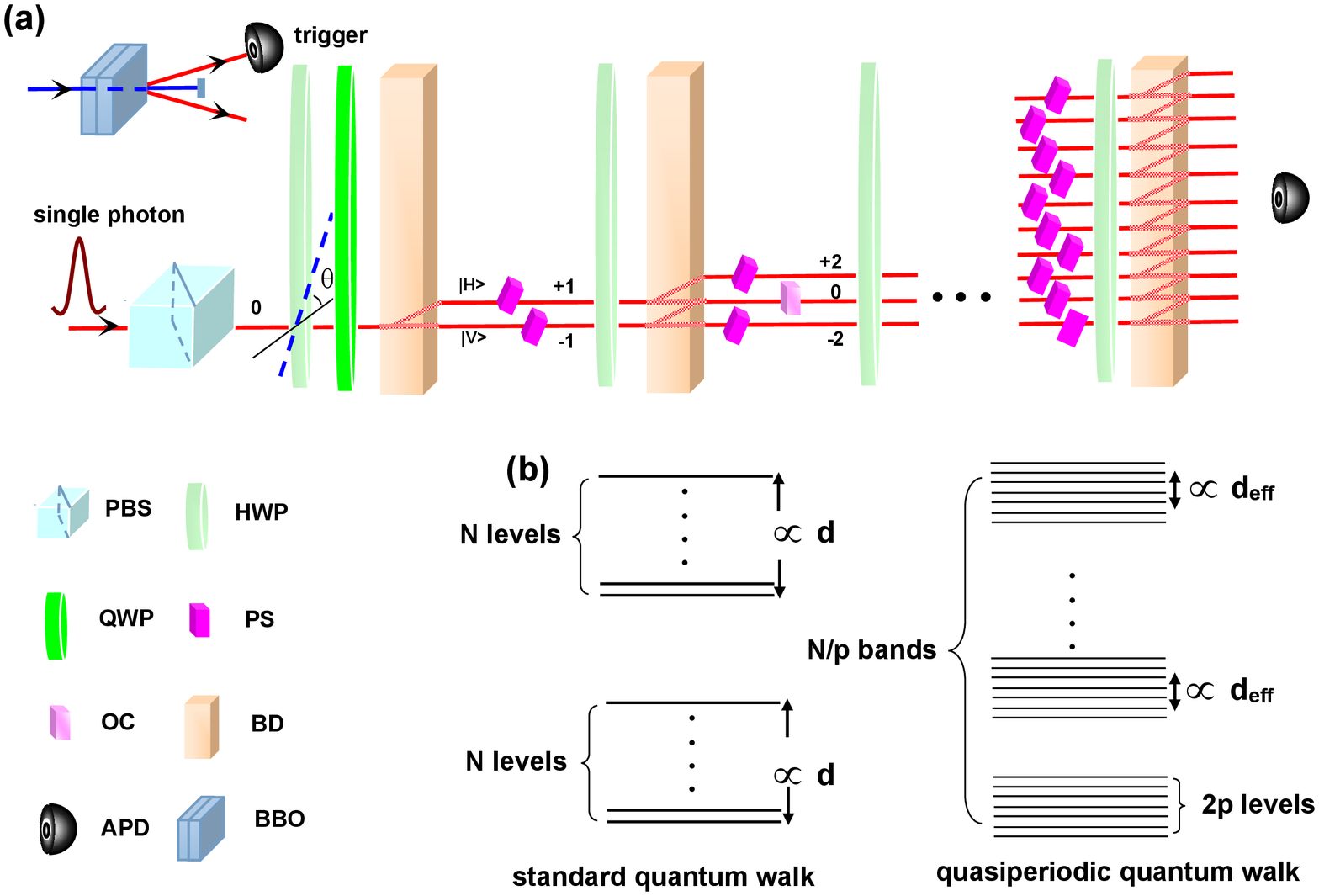}
\caption{
    Experimental scheme for $10$-step QW with position-dependent phase
    function.
    (a)~Single-photons are created via spontaneous parametric down conversion in
    two $\beta$-Barium Borate (BBO) crystals.
    One photon in the pair is detected to herald the other photon,
    which is injected into the optical network.
    Arbitrary initial coin states are prepared by a polarizing beam splitter
    (PBS) and half-wave plates
    (HWP) and quarter-wave plates
    (QWP).
    Phase shifters (PS) are placed in separate transverse spatial modes,
    and optical compensators (OC)
    compensate their resultant temporal delays.
    Coincident detection of trigger and heralded photons at avalanche
    photo-diodes (APDs) yields data for the QW.
    (b)~Quasi-energy spectra for the standard and quasi-periodic QW respectively.}\label{Fig1}
\end{figure}

Optical interferometers comprising birefringent-crystal BDs, wave
plates (WPs), and PSs serve as stable devices for simulating quantum
information processes such as heralded coined QWs. The
interferometer set-up is shown in Fig.~1(a). The $10$-step QW is
implemented with a single HWP, BD, and PSs in each path signifying
one possible site for the walker. The coin state corresponds to the
polarization of the photon, which is produced by SPDC creating
correlated photons pairs at random times. One photon in the pair is
the signal photon, which undergoes the QW, and the other is the
trigger photon, which heralds the presence of the signal photon.
Experimentally QW dynamics is detected by photon coincidence
measurements between the signal and the trigger photon.

Ten BDs, each with length $28.165$mm and a clear aperture of
$33$mm$\times 15$mm, are placed in sequence and need to have their
optical axes mutually aligned. Co-alignment ensures that beams split
by one BD in the sequence yield maximum interference visibility
after passing through a HWP and the next BD in the sequence. In our
experiment, we attain interference visibility of $0.998$ for each
step, i.e., for each pair of sequential BDs.

Output photons from the interferometric network are coupled into a
single-mode optical fibre and subsequently detected by an APD in
coincidence with the trigger photon. We characterize the quality of
the experimental QW by its $1$-norm distance~\cite{BFL+10} from the
simulated QW according to
$\frac{1}{2}\sum_x|P^\text{exp}(x)-P^\text{th}(x)|$; in our case
this distance is $0.085$ after $10$ steps. The distance increases
monotonically with each step number due to a lack of relative phase
control between the multiple interferometers. This limited control
is due to nonplanar optical surfaces. By placing $10$ crystals in
sequence, we are able to achieve~$10$ quantum-walk steps, surpassing
the previous limit of~$7$ quantum-walk steps in a similar
interferometric set-up~\cite{KBF+12}.

The symmetric initial coin state
$(|0\rangle+\text{i}|1\rangle)/\sqrt{2}$ ensures a symmetric walker
position distribution and is prepared by sending the signal photon
through a polarizing beam splitter followed by a HWP and
subsequently by a quarter-wave plate (QWP). A birefringent BD steers
a photon to two possible pathways $3$mm apart in a
polarization-dependent way, which effects the coin-state-dependent
walker translation but without incorporating the position-dependent
phase function.

We introduce a PS in every path between every pair of BDs and
thereby obtain full phase-controlled coin-state-dependent walker
translation. Microscope slides (MSs) with certain effective
thickness are introduced as PSs into the interferometric network
after first completing the alignment described above and ensuring
maximum interference and small distance between simulated and
empirical walker distributions. The MSs are inserted and aligned to
recover the case of zero phase shift $\phi(x)=0$ for all
locations~$x$. This microscope-slide alignment corresponds to each
slide being in the plane perpendicular to the beams in each path.

After achieving this alignment, each MS can be adjusted to an
effective thickness in order impart a controllable phase shift
independent of all other PSs. This effective thickness is achieved
by rotating the slide out of the plane perpendicular to the beams.
Tilting the slide is a viable alternative but is not as stable for
the times required to gather the data.

As an example, we explain how to align the slides for the second
quantum-walk step. In this case there are two PSs because there are
two longitudinal spatial modes after the first step. We rotate the
two PSs separately, then gather photon-count data. These data are
compared to the theoretical simulation. If the data are not
satisfactory with respect to the $1$-norm distance of the walker
distribution, we discard the data, adjust the PSs and repeat. When
the data agree well with the theory, we retain the walker
distribution results and then add the third step of the walker by
adding another three PSs within the three longitudinal spatial modes
respectively, a HWP and BD and repeat this procedure.

Our chief technical innovation in addition to reaching $10$ steps is
the development of fully controllable polarization-independent PSs
that can be inserted into each optical path of a quantum-walk
interferometer. These controllable PSs greatly increase the
versatility of quantum-walk optical-interferometer networks as they
enable generalized QWs with arbitrary phase functions~$\phi(x)$. We
insert PSs into each optical path and adaptively calibrate and
adjust the phases through sequential tests until the alignments
achieve the desired~$\phi(x)$ within acceptable tolerances.

The effectiveness of these controllable PSs is demonstrated by
realizing here quasi-periodic dynamics in a QW for the first time.
Our quasi-periodic evolution has been achieved over $2.5$
quasi-periods, which suffices to see an unambiguous experimental
signature revealing the quasi-period.

Quasi-periodic dynamics arises through symmetry breaking due to the
imposition of the position-dependent phase shift. This
position-dependent phase shift leads to an effective periodic
potential. For~$q$ co-prime to~$p$, the potential wells behave
effectively as a family of~$N/p$ clusters, and we henceforth let~$p$
be even. For the quasi-energy spectrum being the argument of the
eigenvalues of~$U$, the clustering of wells leads to quasi-energy
bands comprising~$2p$ quasi-energy levels each of which only one is
doubly degenerate.

We show the quasi-energy spectra in Fig.~1(b) for $y=q/p$ with the
standard quantum-walk case $q=0$ shown on the left and the $q\neq 0$
case depicted on the right. Since we use two-side coin we have two
bands of Bloch levels, each consisting of~$N$ spectral lines which
are equally spaced for $q=0$, and the width of each band
proportional to the tunneling amplitude $d=\cos\theta$. On the other
hand, for even~$p$ and $q\neq 0$, spectral lines collect into~$p$
Bloch bands with $2p-1$ lines per band and the bandwidth is
proportional to the effective tunneling amplitude
$d_\text{eff}:=d^{p/2}$. Hence~$U^p$ is close to the identity
thereby ensuring a quasi-revival of the initial state after
every~$p$ steps~\cite{WLK+04}.

The quasi-energies $E_{\ell mn}$ of the quantum walker navigating a
lattice with position-dependent phase shifts are given by
$\arg(\lambda_{\ell mn})$ with $\lambda_{\ell mn}$ the eigenvalues
of the unitary operator $U$ and $\ell=0,1,\dots,N/p-1$, $m=0,1$, and
$n=0,1,\dots,p-1$~\cite{WLK+04}. For even~$p$, the eigenvalues
are~\cite{WLK+04}
\begin{equation}
    \lambda_{\ell mn}
        =\text{e}^{2\pi\frac{\text{i}n}{p}}
            \left(2\text{i}r_\ell\left[(-1)^m\sqrt{1-r^2_\ell}-\text{i}r_\ell\right]^{1/p}-1\right),\nonumber
\end{equation}
for $r_\ell=d^{p/2}\sin(\pi p \ell/N)$.

A degeneracy emerges for even~$p$, i.e., for~$E_{\ell mn}$ with
$n\geq p-1$ are same. Therefore, there are a total of~$p$
quasi-energy bands, each consisting of~$2p$ levels with two of them,
namely~$E_{000}$ and~$E_{010}$), being degenerate. For example, in
our case with $p=4$, the resulting system has~4 energy bands, each
band comprising~$7$ quasi-energy levels, and the width of the band
is proportional to $d^{2}$.

For~$d$ small, the bands are narrow, and the system is nearly
harmonic: quasi-periodic dynamics is expected. For~$d$ large,
non-harmonic effects should emerge, and consequently a transition
from quasi-periodic behaviour to ballistic spreading is expected.

We can see the quasi-periodicity arising in our theoretical
simulation and from the experimental data depicted in Fig.~2 for
$y=1/4$, i.e., by setting $q=1$ and $p=4$. In Fig.~2(a), we see the
experimental data for the walker distribution at each step
$t=0,1,2,\dots,10$ with $\theta=\pi/3$. The distribution narrows
almost to the initial spike at $x=0$ for $t=4$ and $t=8$ but has
additional support for $x\neq 0$; i.e., width of the position
distribution has increased. In contrast, for the standard QW
corresponding to any integer-valued $y$, the probability
distribution spreads monotonically with a width proportional to~$t$
as seen in the previous $6$-step standard-quantum-walk
experiment~\cite{BFL+10}.

\begin{figure}
\includegraphics[width=4.2cm]{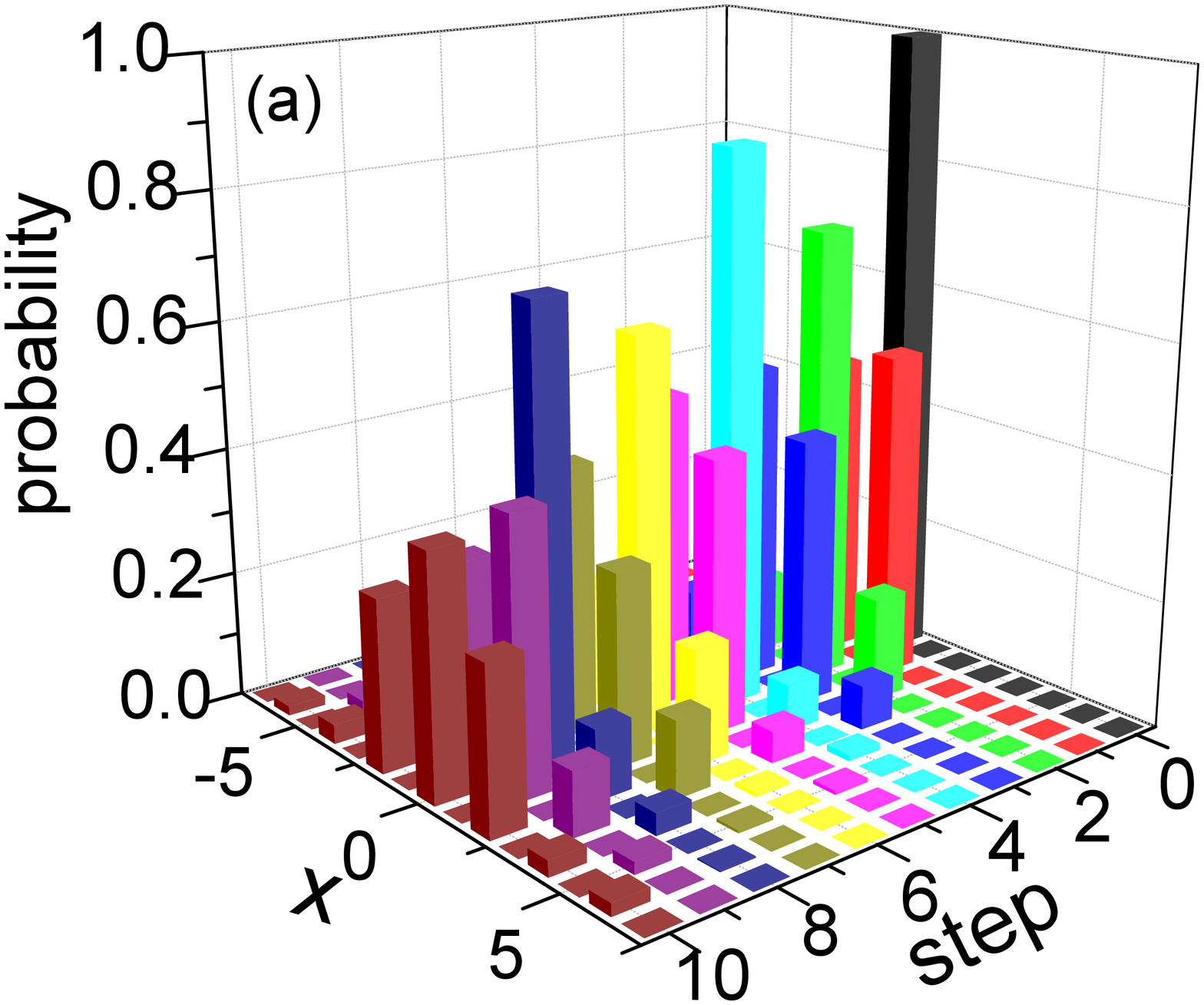}
\includegraphics[width=4.2cm]{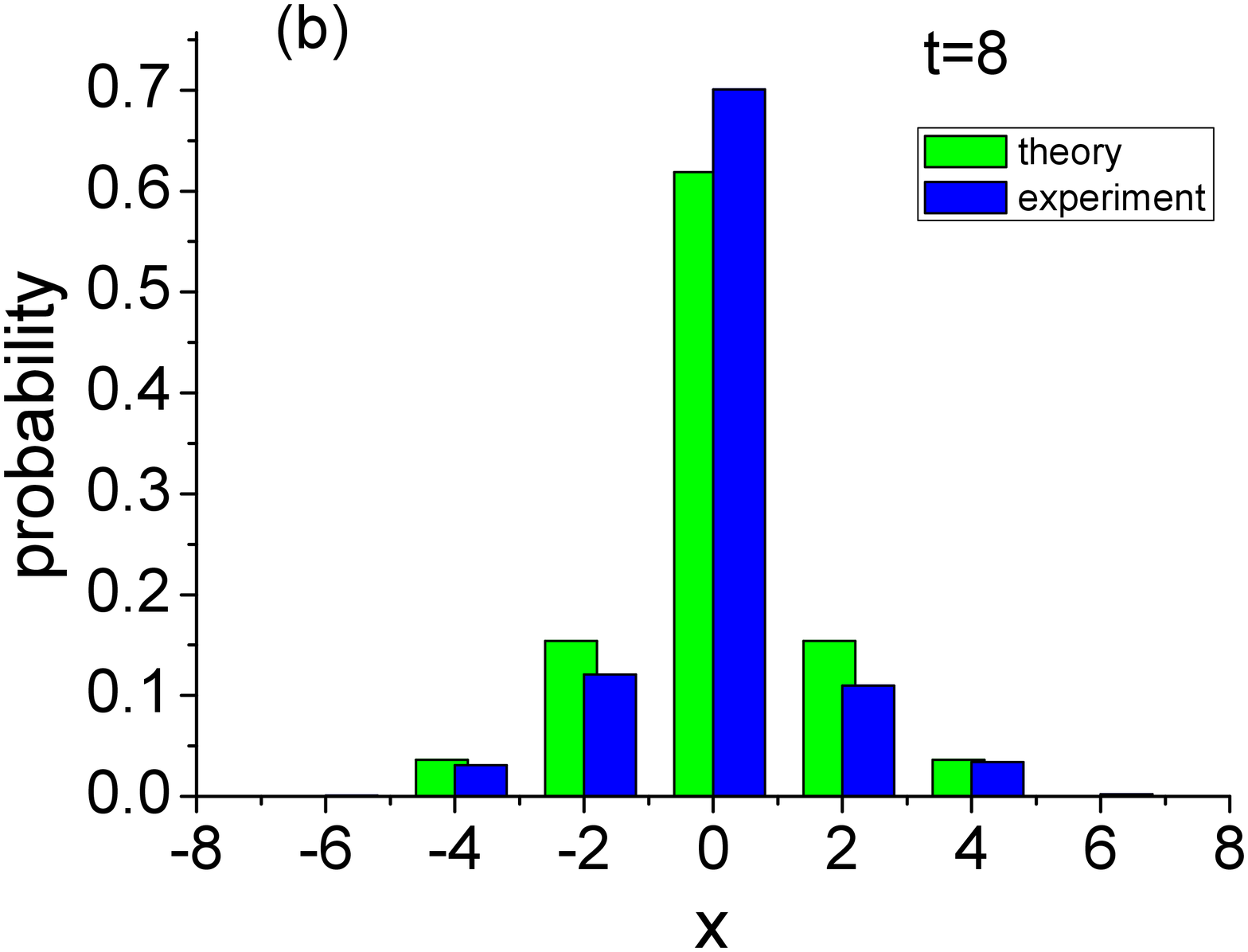}
\includegraphics[width=4.2cm]{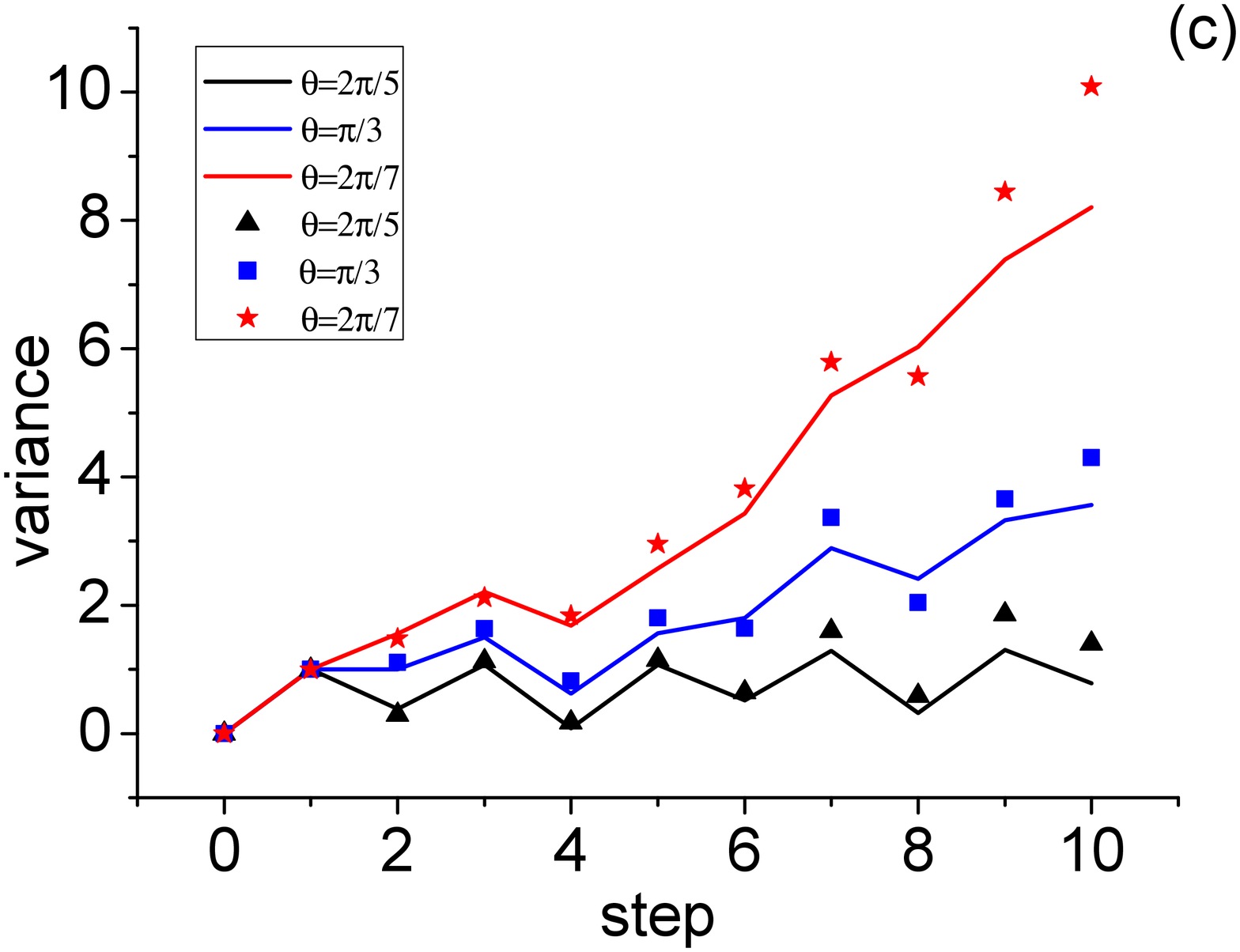}
\includegraphics[width=4.2cm]{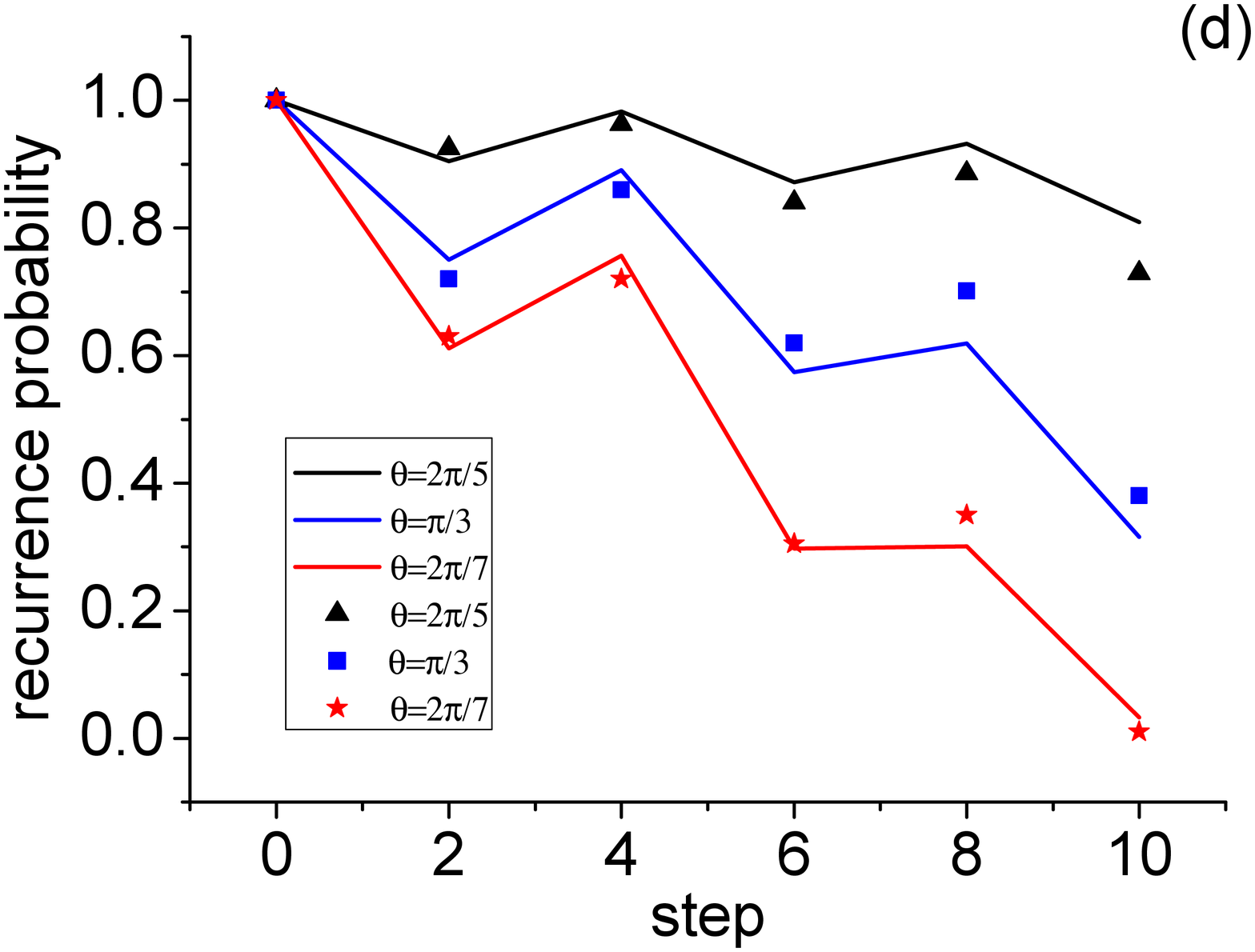}
\caption{Quasi-periodicity of the QW. (a) Experimental position
distributions for successive steps of the $C(\pi/3)$ coined QW with
position-dependent phase function $\phi(x)=2\pi x/4$ and a symmetric
initial coin state up to~10 steps. (b) Probability distribution of
the 8-step QW evolved after two quasi-periods $(t=8)$. Blue and
green bars show experimental data and theoretical predictions
respectively. (c)~Measured position variance and (d)~recurrence
probability of quasi-periodic QW with various coin operations
$C(2\pi/5)$ (black triangles), $C(\pi/3)$ (blue boxes) and
$C(2\pi/7)$ (red stars) for $1$ to $10$ steps, compared to
theoretical predictions (solid lines). Error bars are smaller than
symbol size.}\label{Fig2}
\end{figure}

Theoretical vs.\ experimental results are shown in Fig.~2(b) for
$t=8$, which shows their close agreement and also the degree of
spreading after two periods of quasi-periodic revival of the
position distribution. Figures 2(a,b) suggest a quasi-periodic
revival of the position distribution with period~$\tau=4$, and
Figs.~2(c,d) enable quantitative analysis to determine~$\tau$ for
various values of~$\theta$. The variance and recurrence
probability~$P_t(x=0)$ are shown in Figs.~2(c,d), respectively. Each
plot shows an unambiguous period of $\tau=4$ as expected from our
choice of $p=4$, thereby confirming our numerical simulation and our
experiment.

The coin bias~$\theta$ enables tuning between two extremal modes of
behaviour. One extremal mode is the adiabatic limit corresponding to
a truly periodically varying walker position distribution with a
tendency of trapping at the origin. At the other extreme we have the
diabatic limit of no recurrences with just the standard QW with
ballistic spreading being manifested. The coin bias is a convenient
and effective control of diabaticity vs adiabaticity
because~$\theta$ can set the rate for position spreading without
modifying the nature of the position distribution.

Experimentally the coin bias is controlled by adjusting the angle of
a HWP, and experimental results for the walker's position
distribution are shown in Fig.~3(a) at the first quasi-period $t=4$
for various choices of~$\theta$. Increased narrowing of the
distribution for increased~$\theta$ is commensurate with the
expected transition from diabatic to adiabatic behaviour. The
time-dependent variance for theory and experiment for
various~$\theta$ in Fig.~3(b) further confirm the
diabatic-to-adiabatic transition especially showing the vanishing of
quasi-periodicity in the diabatic limit, yielding instead ballistic
spreading.

Now that we have established that our experiment shows
quasi-periodic dynamics and a full transition from adiabatic to
diabatic behaviour by controlling the coin bias, we proceed to show
the quantum-to-classical transition from quasi-periodic to diffusive
dynamics through controlling decoherence in the optical
interferometer network. This decoherence is controlled by tilting
BDs. A nonzero relative angle $\Delta\theta$ between two successive
BDs, as shown in Fig.~4(a), leads to a spatiotemporal mode mismatch.
This mismatch leads to dephasing between the relative paths and
therefore manifests decoherence of the QW.

We choose $\Delta\theta=9.75^\circ$ to realize full decoherence of
the QW~\cite{BFL+10} and show in Figs.~4(b, c, d) the results for
the walker position distribution and normalized variance in the
coherent and decoherent cases for $t=4$. Thus, we can see the
dramatic transition from quasi-periodic dynamics to ordinary
diffusive dynamics through controlled decoherence on the scale of
one quasi-period.

In summary we have developed a versatile optical quantum network
with single-photon inputs that can simulate QWs with arbitrary
position-dependent phase shifts. We show quasi-periodic dynamics of
a quantum walker and clear signatures of adiabatic vs diabatic
behaviour as well as quasi-periodic-to-diffusive dynamics by
controlling decoherence. Our results show a new realm of
quantum-walk phenomenon which is especially interesting in the
context of quantum chaos and Bloch oscillations and our new
interferometer phase-shift control provides a valuable new tool for
exploring QWs with various potentials.

\begin{figure}
\includegraphics[width=4.2cm]{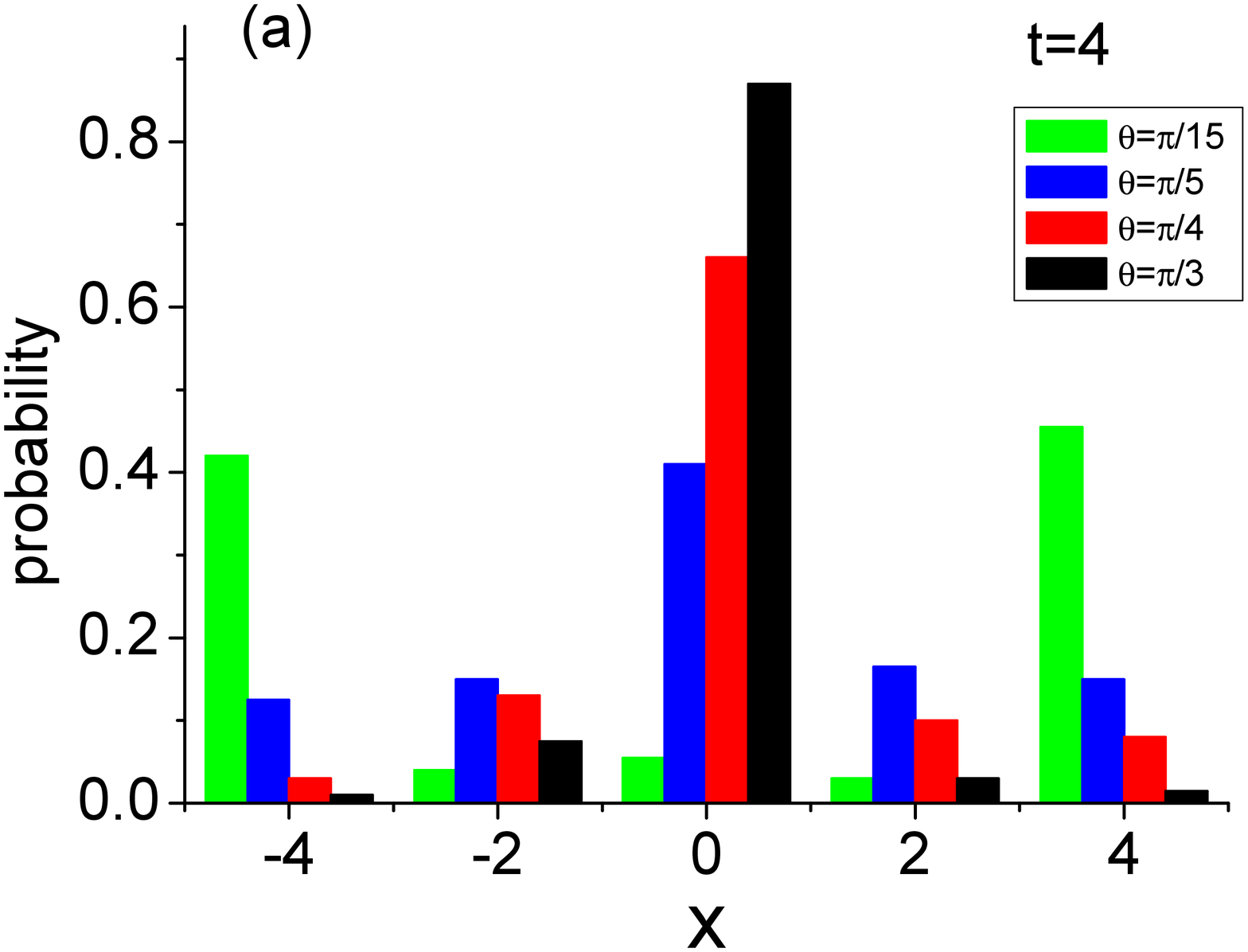}
\includegraphics[width=4.2cm]{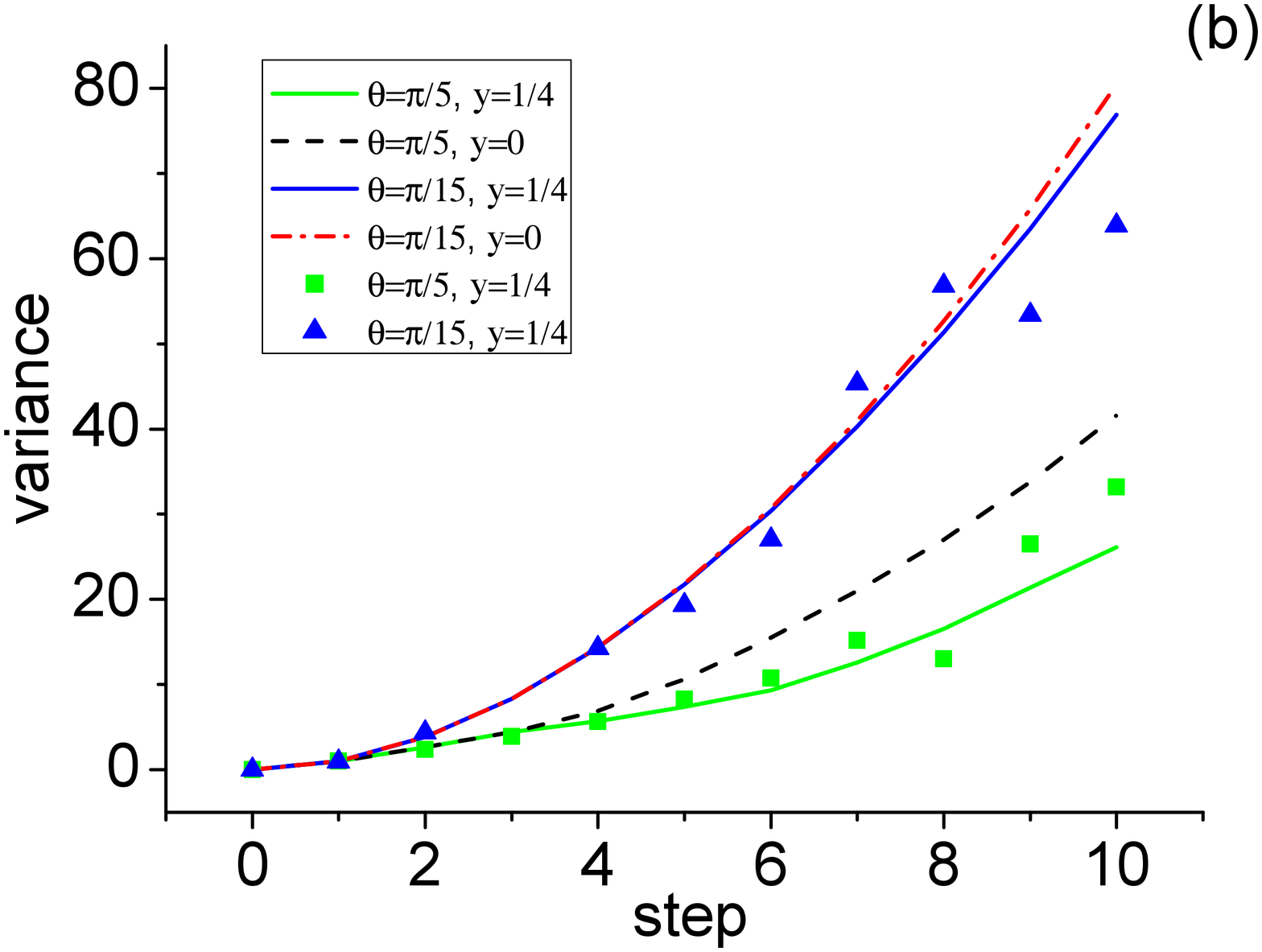}
\caption{
    Adiabatic to diabatic transition of the QW.
    Transition for various coin-flip biases, position-dependent phase function $\phi(x)=2\pi
x/4$ and a symmetric initial coin state. (a)~Experimental position
distributions after the first quasi-period $\tau$. (b)~Measured and
predicted variances for $10$ steps compared to a standard QW
simulation with $\phi(x)\equiv 0$. Error bars are smaller than
symbol size.
    }
\label{Fig3}
\end{figure}

\begin{figure}
\includegraphics[width=4.2cm]{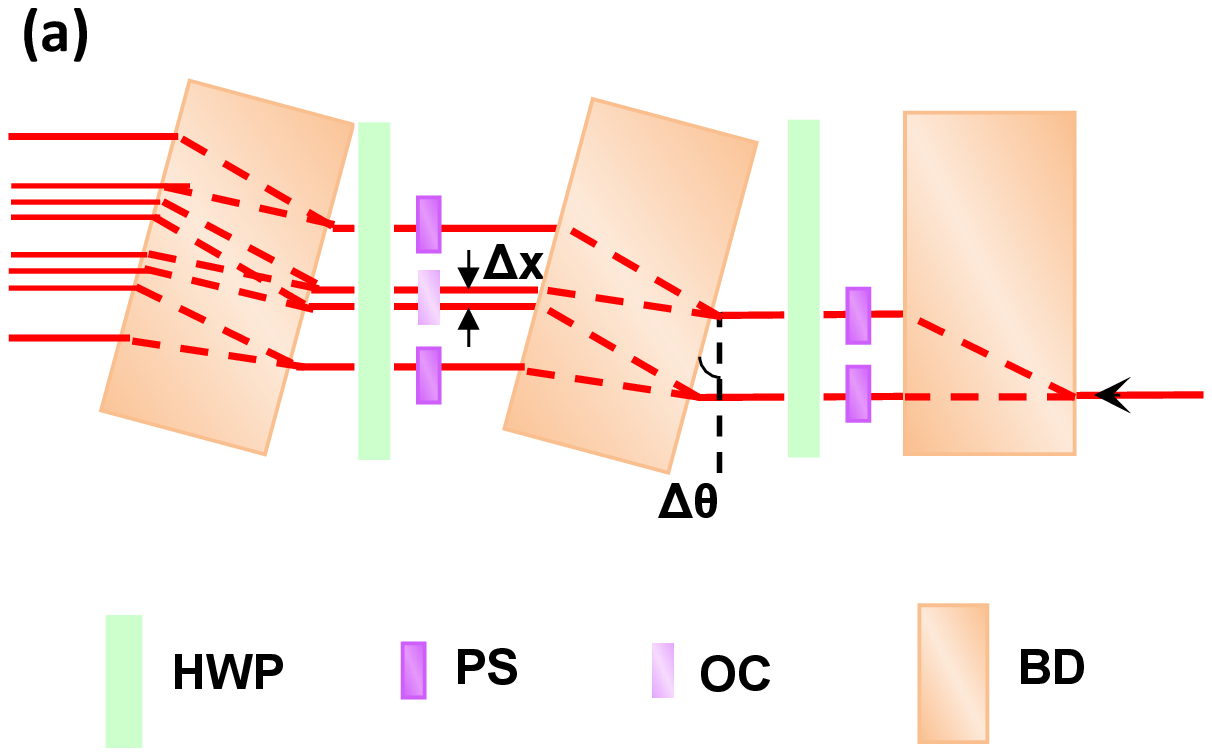}
\includegraphics[width=4.2cm]{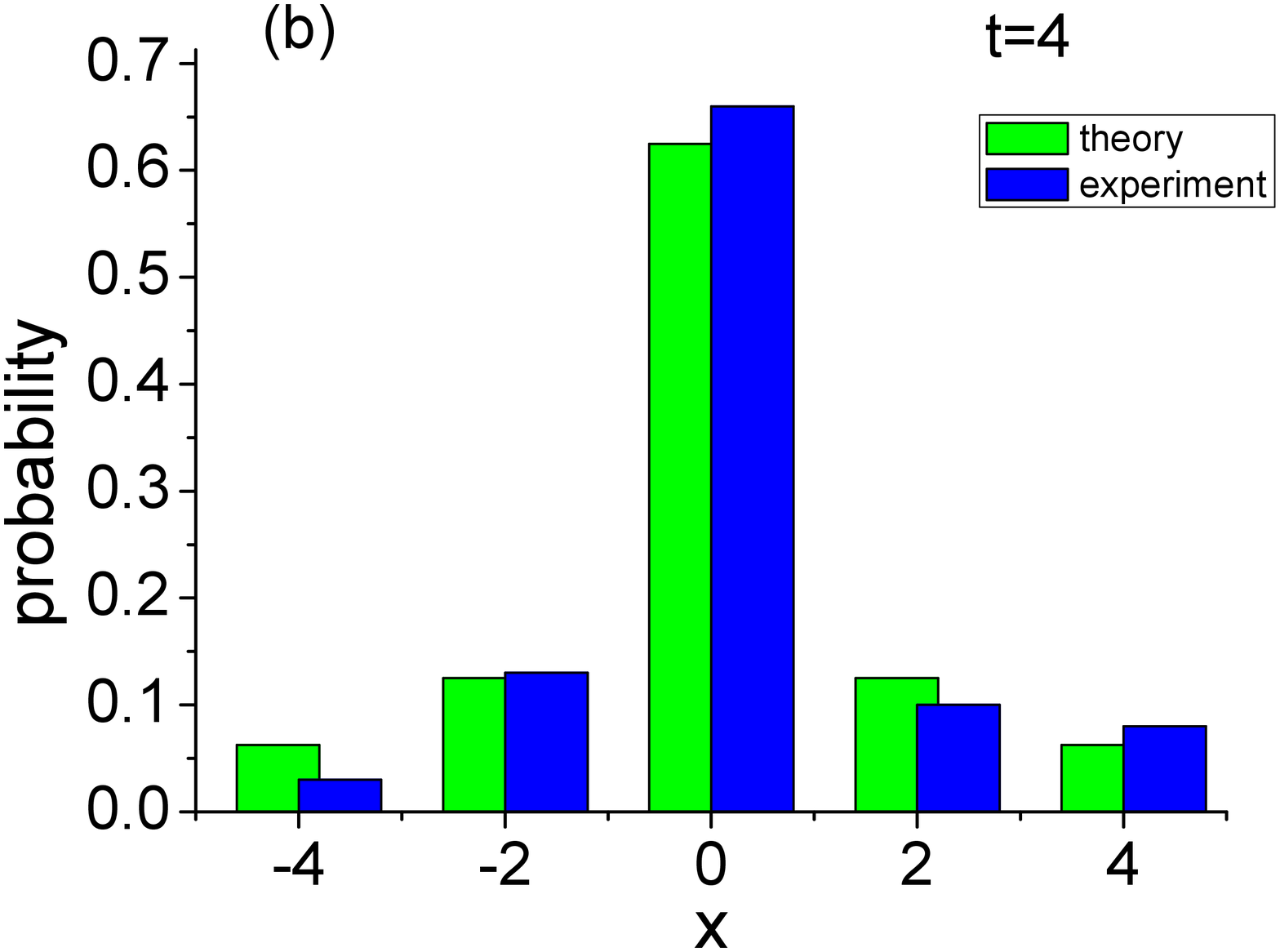}
\includegraphics[width=4.2cm]{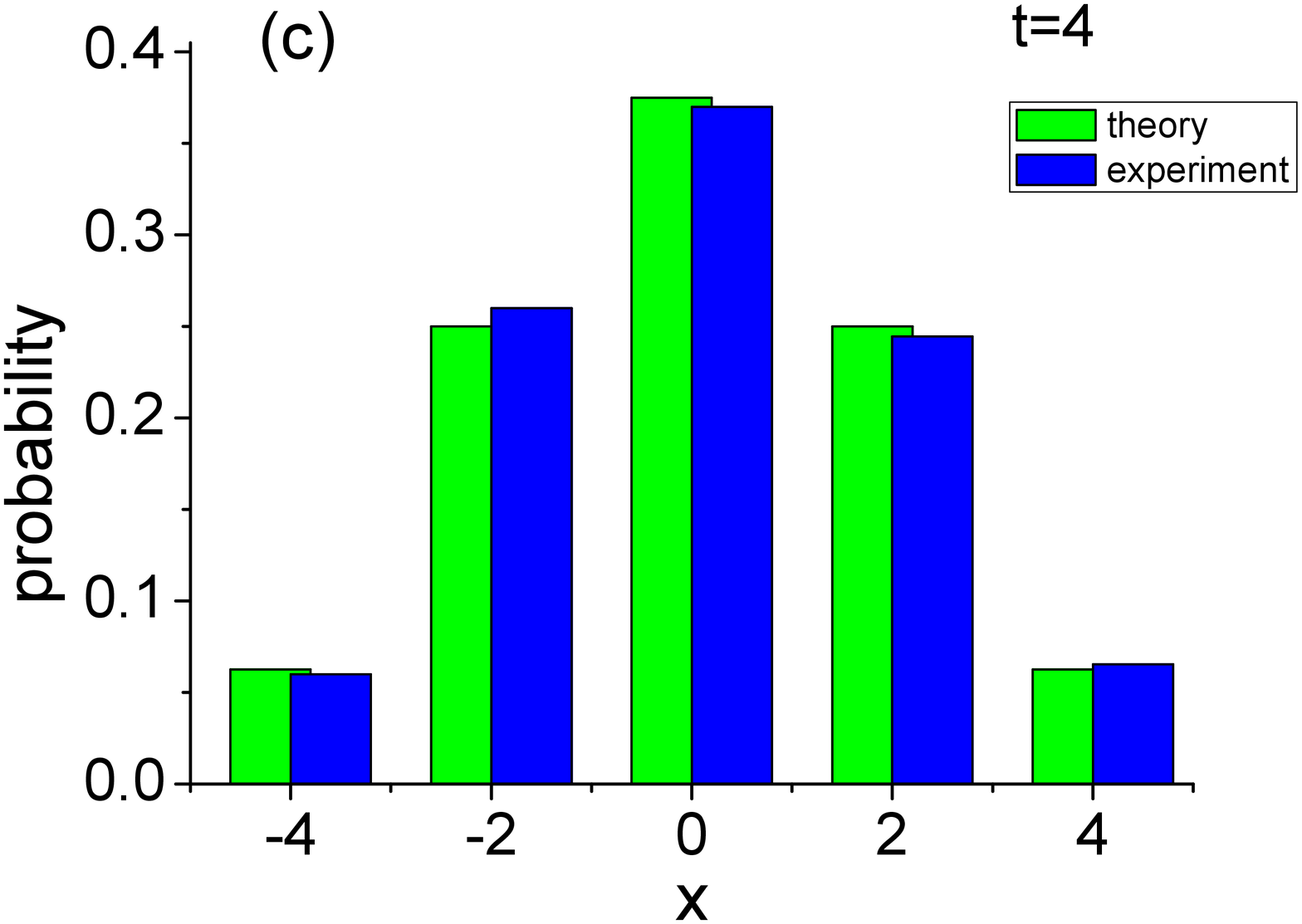}
\includegraphics[width=4.2cm]{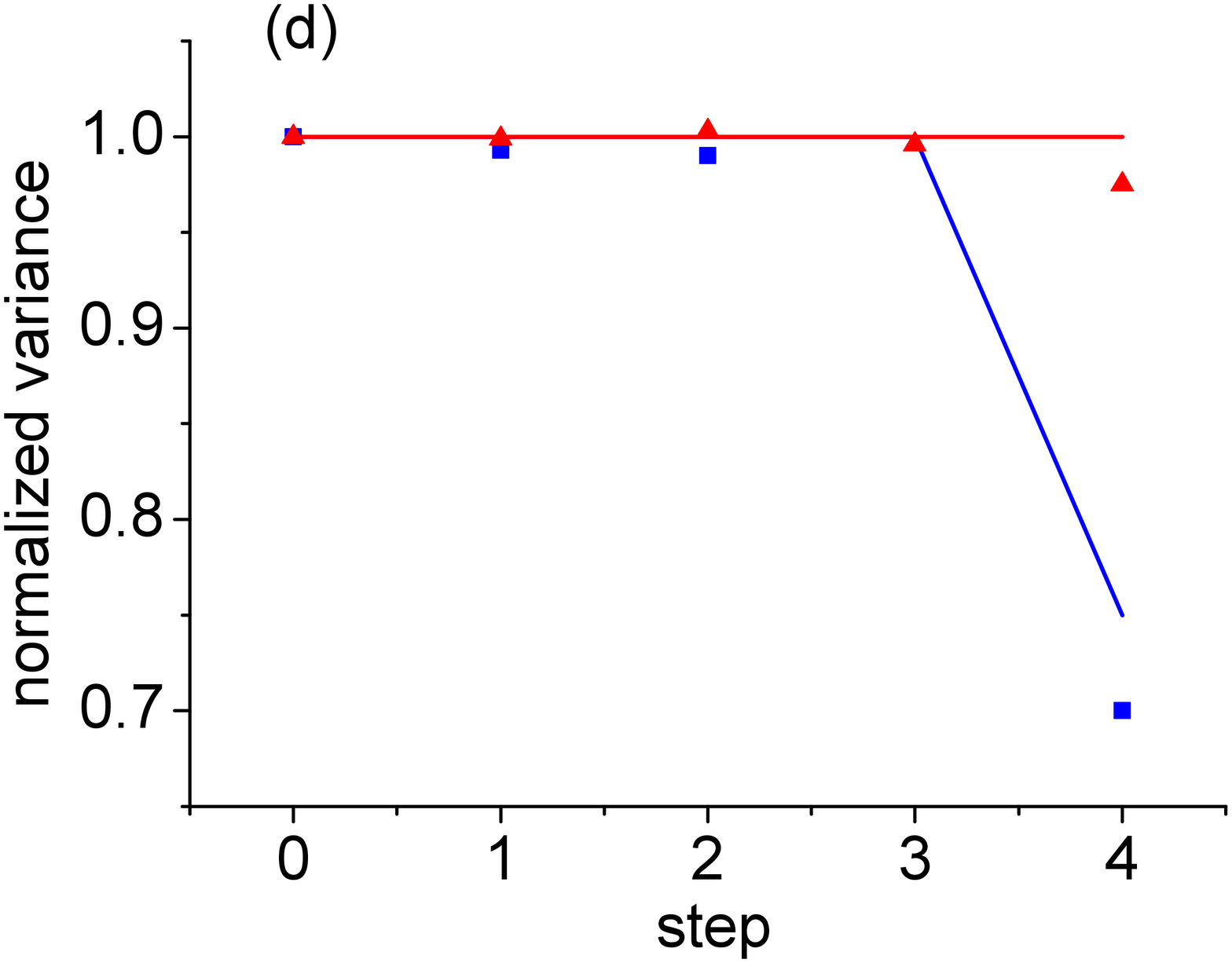}
\caption{Transition from quasi-periodic QW to fully decohered QW.
(a) A relative angle $\Delta\theta$ between two beam displacers
reduces the recombined photon's spatial mode overlap, which
introduces controllable dephasing. (b) Measured position
distribution at the first quasi-period with coin bias
$\theta=\pi/4$, position-dependent phase function $\phi(x)=2\pi x/4$
and a symmetric initial coin state. (c) Position distribution of a
fully decohered QW. Blue and green bars show experimental data and
theoretical predictions respectively. (d)~Normalized position
variances of quasi-periodic QW (blue boxes) and of fully decohered
QW (red triangles) up to~$4$ steps, compared to theoretical
predictions (solid lines). Error bars are smaller than symbol
size.}\label{Fig4}
\end{figure}

\acknowledgements We acknowledge experimental guidance from Xiao-Ye
Xu, Kai Sun, Jin-Shi Xu, Chuan-Feng Li and Guang-Can Guo. This work
has been supported by NSFC under 11004029 and 11174052, NSFJS under
BK2010422, 973 Program under 2011CB921203, the Open Fund from the
State Key Laboratory of Precision Spectroscopy of East China Normal
University, the 1000 Talent Program of China, the Canadian Institute
for Advanced Research, and Alberta Innovates Technology Futures.

\end{document}